# Benchmarking and Validation of Sub-mW 30GHz VG-LNAs in 22nm FDSOI CMOS for 5G/6G Phased-Array Receivers

Domenico Zito, AGH University of Science and Technology, 30-059 Krakow, Poland; domenico.zito@icloud.com

Michele Spasaro, Aarhus University, 8200 Aarhus, Denmark

*Abstract*— Next-generation (5G/6G) wireless systems demand low-power mm-wave phased-array ICs. Variable-gain LNAs (VG-LNAs) are key building blocks enabling hardware complexity reduction, performance enhancement and functionality extension. This paper reports a performance benchmarking of two low-power 30GHz VG-LNAs for phased-array ICs, which provide a 7.5dB gain control for 18dB Taylor taper in a 30GHz 8x8 antenna array, for a comprehensive validation of the new class of VG-LNAs and its design methodology. In particular, this paper reports a second and implementation (VG-LNA2) with a reduced number (four) of gain-control back-gate voltages and super-low-Vt MOSFETs, with respect to the previous first implementation (VG-LNA1) with six gain-control back-gate voltages and regular-Vt MOSFETs, both in the same 22nm FDSOI CMOS technology. The results show that VG-LNA2 exhibits performance comparable to those of VG-LNA1, with a slightly lower power consumption. Overall, the performance benchmarking shows that the design methodology adopted for the new class of VG-LNAs leads to record low-power consumption and small form factor solutions reaching the targeted performances, regardless of the arrangements of the back-gate voltages for gain control and transistor sets, resulting in a comprehensive validation of the innovative design features and effective design methodology.

*Keywords—Variable gain low noise amplifiers, phased arrays, 22nm FDSOI CMOS.*

## I. INTRODUCTION

The 5G New Radios (NRs) with operations in the Frequency Range 2 (FR2), i.e., 24.25-52.6 GHz, and the on-going standardization of the next-generation cellular systems are pushing the boundaries of the design and implementation of phased-array ICs for 5G/6G wireless communications. The first reported 28GHz IC for 5G [1] consists of 32 TR/RX (TRX) elements, each with a TX/RX switch, a 12dB gain low noise amplifier (LNA) with noise figure (NF) of 6 dB, which comprises the losses due to the switch and off-mode power amplifier (PA), and a variable-gain amplifier (VGA) for beam shaping, as illustrated in Fig. 1.

Some wireless applications can take advantage of switchless TRX architectures (see Fig. 1), which are typically implemented as asymmetrical structures (N ≠ M) to accomplish the communication needs and reduce the better hardware complexity. The removal of lossy TX/RX switches is key to relax the gain and noise requirements for the LNAs, favoring low-power designs.

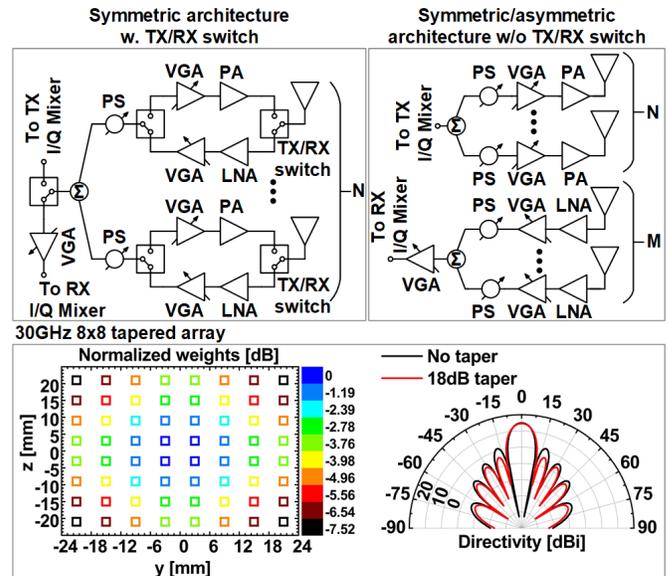

Fig. 1. Phased-array TX/RX architectures for phased-array ICs with RF phase shifting (top). Gain control and beamforming in 30GHz 8×8 3GPP TR 38.901 compliant-antenna array (6mm pitch) with 18dB Taylor window taper (bottom).

Furthermore, the LNAs in the RX elements can be replaced with variable-gain LNAs (VG-LNAs) to further reduce the overall power consumption (PC). VG-LNAs are also key building blocks in phase-array front-ends with digital beamforming [2]. Thereby, compact low-power VG-LNAs are key for the implementation of cost-effective energy-efficient low-power phased-array ICs for future-generation (5G/6G) wireless transceivers.

Among the prior works, [2-6] reported low-power VG-LNAs for 5G/6G phased-array ICs, and wideband VG-LNAs [5] for high data-rate long-haul wireless communications. Reported gain control techniques include current steering [2], tunable resistive loads and bias current control [5], and bias current control with phase compensation based on the complementary dependence of p- and n-MOSFET capacitances. LNAs [2-5] consume a few tens of milliwatts and occupy silicon areas larger than 0.1 mm$^2$, owing to lossy and large passive components such as spiral inductors, transformers, transmission lines (TLs), as well as the adoption of differential topologies [3-5]. In [6], we have addressed the feasibility of a sub-mW mm-wave VG-LNA for 5G/6G for low-power phased-array receivers for the imple-

This work was supported in part by the Poul Due Jensen Foundation (PDJF) and in part by the European Commission through the European H2020 FET Open project IQubits (GA N. 829005).

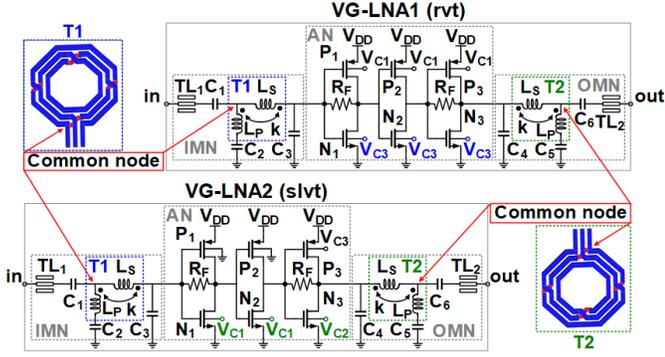

Fig. 2. Schematics of the two low-power VG-LNAs: VG-LNA1 (top); VG-LNA2 (bottom). Dashed boxes identify the custom-designed transformers (T1, T2) with their layouts, active network (AN), input matching network (IMN) and output matching network (OMN).

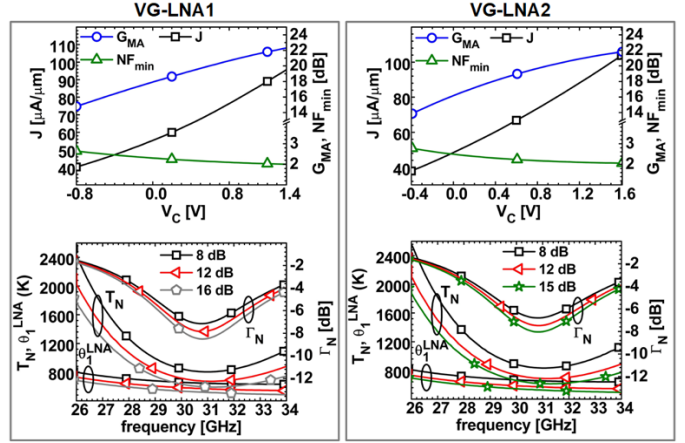

Fig. 3. Post-layout simulation results: Transistor current density (J), and maximum available gain ($G_{MA}$) and $NF_{min}$ of the AN at 30 GHz, vs. control voltage ($V_C$); $T_N$, $\theta_1^{LNA}$ and $\Gamma_N$ [17] of the VG-LNAs vs. frequency, for low, medium (nominal) and high gain states.

mentation of a 18dB Taylor window taper with three approximately equal side lobes in a planar 8x8 antenna array operating at 30 GHz, as illustrated in Fig. 1, and explained in [6]. Therein [6], the first VG-LNA of a new class of VG-LNAs, i.e. VG-LNA1, has been implemented with regular-Vt (rvt) MOSFETs and the measurement results have shown the effectiveness of the design approach, which has allowed reaching record low-power consumption performance. Further details about the circuit design are reported in [7].

This paper reports a second and new implementation of such a new class of sub-mW VG-LNAs, with the circuit design featuring super-low-Vt (slvt) MOSFETs and gain control reduced to only four back-gate voltages, as shown in Fig. 2. Then, the results obtained for the two VG-LNA design implementations with the different features as above, both carried within the same 22nm fully-depleted silicon-on-insulator (FDSOI) CMOS technology, are compared for a comprehensive performance benchmarking and overall validation of the circuit topology with the own inherent design features and design methodology. In order to carry out the performance benchmarking and design methodology validation in the frame of the targeted application in future 5G/6G phased-array receivers, the main results reported in [6] for the design implementation with regular-Vt MOSFETs are summarized here to be compared with those of the second and new implementation, which is reported here for the first time.

The paper is organized as follows. Section II summarizes the main features of the design implementations of the two VG-LNAs. Section III reports the measurement results and benchmarking. Finally, in Section IV, the conclusions are drawn.

## II. DESIGN OF THE TWO VG-LNAS

Both the VG-LNAs, VG-LNA1 implemented with regular-Vt (rvt) and VG-LNA2 with super-low-Vt (slvt) MOSFETs, have been designed to implement 18dB Taylor window taper with three approximately equal side lobes in a planar 8x8 antenna array operating at 30 GHz [6].

Fig. 1 shows the normalized weights and the directivity of the antenna array without and with tapering, obtained from simulations considering antenna elements compliant with 5G NR technical report 3GPP TR 38.901. The results show that a side lobe suppression of 18 dB is achieved with a 7.5dB gain control range. This result is also coherent with the experimental tests reported in [1]. Therefore, the VG-LNA1 and VG-LNA2 have been designed to exhibit gain control ranges of 7.5 dB, with sub-mW PC for all gain states, and with a peak gain of about 16 dB in the high gain state, in compliance with the tapered phased-array requirements. A NF of about 6 dB was considered as a reference value aligned with other designs (e.g. [1]). Further considerations about noise are reported in [6].

Fig. 2 reports the schematics of the two VG-LNAs. The back-gate voltages $V_{C1}$-$V_{C3}$ are used as gain control voltages provided through DC pads. In both VG-LNAs, the gain is varied by applying a single control voltage ($V_C$) to the back-gates of all n-MOSFETs, according to the following scheme: For VG-LNA1, $V_{C1} = 0$ V and $V_{C3} = V_C$; for VG-LNA2, $V_{C3} = 0$ V and $V_{C1} = V_{C2} = V_C$. The active networks (ANs) exploit complementary current reuse and consist of three gain stages: A self-biased inverting amplifier (1st stage) and a Cherry-Hooper amplifier (2nd and 3rd stage), to achieve a nominal gain of 12 dB with a power consumption of about 0.6 mW.

Unlike the well-established design approach for microwave and mm-wave ICs (e.g., [8-15]), the amplifier stages of the ANs do not include any spiral inductors/transformers for peaking and inter-stage matching networks (MNs), which are typically implemented with the thick top-metal layers in order to secure a high quality (Q) factor. Indeed, the traditional design approach requires connecting the transistors (laying at the bottom) to the top metal layers all through the back-end-of-line (BEOL), and this connection introduces substantial losses that are typically compensated by increasing the power consumption. Therefore, the new design approach with no spiral inductors/transformers into the ANs leads to a paradigm shift enabling low-power implementations with very compact layouts and reduced parasitics, which allow taking better advantage of inherent performances of the ultra-scaled MOSFETs, owing to the drastic reduction of lossy interconnects from the bottom to the top metal layers, whose effects are predominant in low-power designs at high frequencies [16, 17]. Moreover, as a positive side effect, in principle such an approach with reduced parasitics allows more broadband operations. However, the effectiveness of such an approach decreases as the operating frequency increases, as a consequence of the inherent parasitic capacitance of the transistors. The experimental validation of the VG-LNA1 [6] has demonstrated that such an approach provides good results at 30 GHz, i.e., at the frontier between microwave and mm-wave frequencies. A pilot implementation for quantum computing applications at cryogenic temperatures (2 K) has also shown that such a design approach can reach adequate performance for operations at 60 GHz [18].

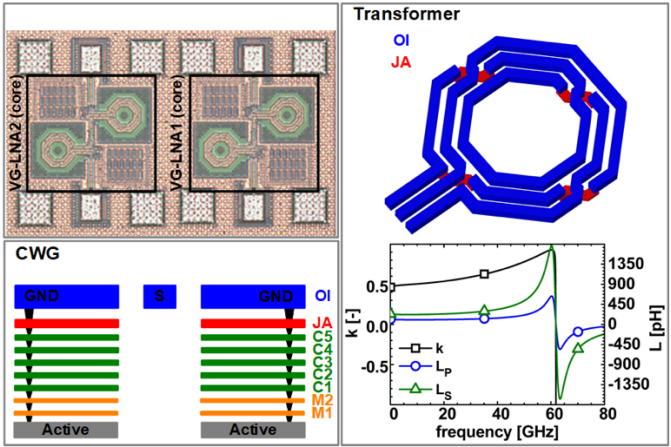

Fig. 4. Test-chip micrograph of VG-LNA1 and VG-LNA2 (left). 3D layout view of the transformers and EM simulation results (right); sketch (excluding fillers) of the coplanar waveguides to the RF pads (bottom left).

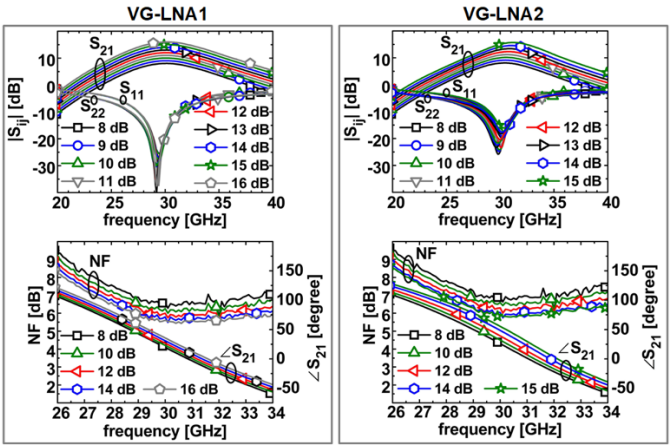

Fig. 5. Measured S-parameters and NF of the two VG-LNAs vs. frequency, for all gain states.

All transistors have a gate width (W) of 3.9 μm. $V_{DD}$ of VG-LNA1 and VG-LNA2 amount to 800 mV and 680 mV, respectively. The AN exhibits an input impedance close to 200-j400 Ω at 30 GHz for both VG-LNAs in the nominal gain state of 12 dB. The input impedance is then transformed to 50 Ω through the input matching network (IMN), consisting of the cascade of two L- networks: $C_1$-$L_P$ and $L_S$-$C_3$, with the magnetic coupling to improve the equivalent Q-factor of the two spiral inductors and reduce the area on chip. $L_S$-$C_3$ allows the impedance transformation of the input impedance of the AN in an intermediate impedance $Z_{IM}$, which is used as a free parameter to facilitate the integrated implementation of the IMN and to improve the noise performance of the VG-LNA [17]. The coupling factor k is a further design parameter used to minimize the cascade noise [17]. Similar considerations can apply to the output matching network (OMN), in which the cascade gain is optimized instead of the cascade noise.

Fig. 3 reports the results of the post-layout simulations (PLS). In particular, Fig. 3 (top) shows the maximum available gain ($G_{MA}$) and minimum NF ($NF_{min}$) of the ANs at 30 GHz, and transistor current density (J) plotted as a function of the control voltage $V_C$. For VG-LNA1, J increases from 41 to 94 μA/μm as $V_C$ varies from –0.75 to 1.35 V. For VG-LNA2, J increases from 42 to 104 μA/μm as $V_C$ varies from -0.25 to 1.6 V. For both VG-LNAs, the increase of J leads to an increase of $G_{MA}$ from 15 to 22 dB, and a reduction of the $NF_{min}$ from 2.6 to 2.0 dB as J approaches the optimum-noise current density of about 0.1-0.15mA/μm. Fig. 3 (bottom) reports also the equivalent noise temperature ($T_N$) and the lower bound $\theta_1^{LNA}$ [17]. Also, the same figure reports the magnitude of the optimum-noise impedance matching coefficient $\Gamma_N$ [17]. $|\Gamma_N|$ has a minimum at about 31 GHz, leading to a $T_N$ approaching the theoretical lower bound $\theta_1^{LNA}$ [17]. As J increases, $|\Gamma_N|$ decreases and this further improves $T_N$ for the higher gains.

Fig. 4 (top left) reports the die micrograph. Both the VG-LNAs have a core area of 0.20×0.22 mm² and an overall area footprint on die of 0.32×0.26 mm² including RF pads. Fig. 4 (right) illustrates the layouts of the coupled inductors, i.e., transformers T1 and T2 implemented with the two topmost metal layers (OI and JA). The electromagnetic (EM) simulations show that $L_P$, $L_S$, and k amount to 119 pH, 267 pH, and 0.59, respectively. Each MN includes a short (50 μm) 50Ω coplanar waveguide to the test RF pads.

III. MEASUREMENTS

The VG-LNAs were measured on die with the Keysight PNA-X N5245A. Power calibrations were carried out with the power meter N1914A and power sensor N8487A by Keysight.

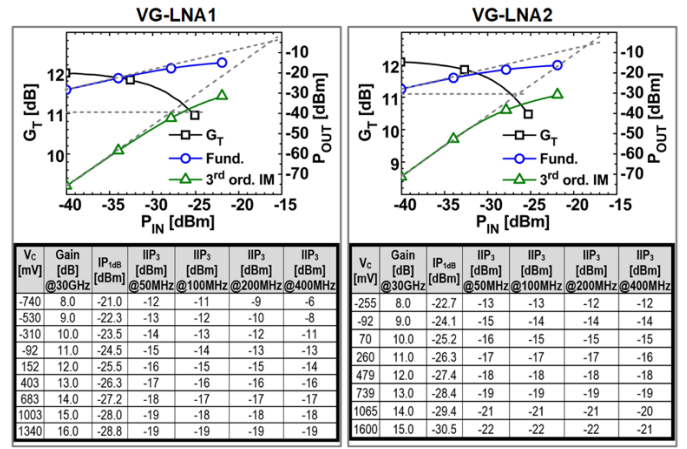

Fig. 6. Measured linearity performances of the two VG-LNAs in the nominal (12dB) gain state (top): Transducer gain ($G_T$) at 30 GHz and average power of the fundamentals and 3rd order intermodulation products (IM) for input tones at 30 and 30.05 GHz (top). Measured $IP_{1dB}$ and $IIP_3$ for all gain states (bottom) and two-tone tests (Δf = 50-400 MHz).

Table 1. Comparison with prior-art VG-LNAs operating in 5G NR FR2.

| Work | VG-LNA2 | | VG-LNA1 [6] | | [2][1] | | [3] | | [4] | | [5] | | [19] | |
|---|---|---|---|---|---|---|---|---|---|---|---|---|---|---|
| CMOS Tech. [nm] | 22 (FDSOI) | | 22 (FDSOI) | | 22 (FDSOI) | | 40 (Bulk) | | 65 (Bulk) | | 90 (Bulk) | | 65(Bulk) | |
| Peak gain [dB] | 15.7 | 8.1 | 16 | 8.0 | 24.2 | 16.2[2] | 27.1 | 18.4 | 20.8 | 10.2 | 21.4 | 11.6 | 11.4 | -5 |
| $f_c$ [GHz] | 31.2 | 30.4 | 30.1 | 30.1 | 30.4[5] | 30.4[5] | 27.1 | 27.8 | 30.4[2] | 30.4[2] | 37 | 37[2] | 28 | 28 |
| $BW_{3dB}$ [GHz] | 5.7 | 6.4 | 7.1 | 7.2 | 23.7 | 23.7[2] | 7.4 | 9.3 | 4 | 4[2] | 11.3[4] | 11.3[2] | 11.5 | - |
| Min. NF [dB] | 5.9 | 6.9 | 5.5 | 6.3 | 2.4 | 3.0 | 3.3 | 3.4 | 3.7 | - | >4.7 | - | >4.7 | - |
| $IP_{1dB}$ [dBm] | -30.5[3] | -22.7[3] | -28.8[3] | -21.0[3] | -25 | -19 | -21.6 | -13.4 | -20.4 | - | -25.1[5] | -22[2] | - | - |
| ∠$S_{21}$ variation* [°] | 29.5 | | 16.7 | | - | | 18 | | 8 | | 7.2 | | - | |
| $P_C$ [mW] | 0.91 | 0.40 | 0.97 | 0.41 | 16 | 16 | 31.4 | 21.5 | 26.7 | 16.5 | 17.9 | 17.9 | 2.16 | - |
| Core area [mm$^2$] | 0.044 | | 0.044 | | - | | 0.26 | | 0.20 | | 0.45[6] | | 0.13 | |

[1]Simulation results; [2]deduced from plots; [3]measured at 30 GHz; [4]dual band (26-30.5 GHz and 33.8-40.6 GHz) [5]calculated from available data; [6]including pads; *max variation in 3dB band.

Fig. 5 reports the S-parameters $S_{21}$ (mag. and phase), $|S_{11}|$ and $|S_{22}|$, and NF of the two VG-LNAs for all gain settings. In the high-gain state, VG-LNA1 exhibits a peak gain of 16 dB at 30.1 GHz and a minimum NF of 5.5 dB with a PC of 0.97 mW; VG-LNA2 exhibits a peak gain of 15.7 dB at 31.2 GHz and a minimum NF of 5.9 dB with a PC of 0.91 mW. In the low-gain state, the PC of VG-LNA1 and VG-LNA2 amounts to 0.41 mW and 0.40 mW, respectively. The $S_{12}$ of VG-LNA1 and VG-LNA2 is lower than -33 dB and -22 dB, respectively, for all gain states. These measurement results confirm the simulation results. The two VG-LNAs exhibit comparable performances compliant with the reference requirements. However, the VG-LNA2 exhibits a lower PC (up to 60 μW less, i.e., ~7% less), which corresponds to a further new record low power consumption.

Fig. 6 reports the measured input-referred $P_{1dB}$ ($IP_{1dB}$) and $IP_3$ ($IIP_3$) of the VG-LNAs for all gain settings. Two-tone tests were carried out with two tones at $f_1$=30 GHz and $f_2$=30 GHz+Δf, with Δf=50, 100, 200, and 400 MHz, as per 5G channelization. The plots show the results for the VG-LNAs in the nominal 12dB gain state with Δf=50 MHz; the tables, show the results for all gain states and two-tone tests. Despite the sub-mW power consumption, both the VG-LNAs exhibit comparable linearity performances, aligned with the application requirements (approx. -30 dBm).

Overall, both the VG-LNAs exhibit adequate performances compliant with the targeted switchless tapered phased-array TRX architecture, effective gain control, and record power consumption below 1 mW and area occupancy. Despite the sub-mW power consumption, the two VG-LNAs exhibits acceptable linearity performance close to the prior-art works, e.g. [2], [5], which exhibit a power consumption higher than one order of magnitude. Also, it is worth emphasizing that, unlike other works in the prior art, e.g. [2], [5], the power consumption of the two VG-LNAs reduces as the gain reduces, and this is a distinctive and very effective feature enabling energy-efficient low-power phased-arrays. Last, it is worth observing that, irrespective of the gain states, the results show that 16-to-32 units of VG-LNA1, or VG-LNA2, consume less than one unit of the prior-art VG-LNAs.

## IV. CONCLUSIONS

This paper reports the implementations of two low-power 30GHz VG-LNAs for 5G/6G phased-array ICs, capable of providing a gain control of 7.5 dB for 18dB Taylor taper in a 30GHz 8x8 antenna array, with the objective to carry out a performance benchmarking and a comprehensive validation of the new class of VG-LNAs. In particular, the paper reports two different implementations of the same circuit designed with the same methodology, but with a different arrangements of the back-gate voltages for the gain control: six in the implementation reported previously, and four in the implementation reported here for the first time; and, different transistor sets: regular-Vt for the design reported previously, and super-low-Vt for the design reported here for the first time, both transistor sets available within the same 22nm FDSOI CMOS technology. Despite the different design implementations, the measurement results show that both the VG-LNAs reach the performance requirements with a peak gain of about 16 dB in the high gain state, consume less than 1 mW, and occupy a very compact area of 0.20×0.22 mm$^2$.

In particular, the results show that the second and new implementation with super-low-Vt MOSFETs and a reduced number (four) of back-gate control voltages, reported here for the first time, allows reaching performances comparable to those achieved previously with the regular-Vt MOSFETs, with a lower power consumption (~7% less), which sets a further new record low-power consumption. Therefore, from the performance benchmarking carried out in the frame of the targeted applications for 5G/6G low-power tapered phased-array receivers, the results obtained for the two different implementations provide a comprehensive validation of the effectiveness of the new class of VG-LNAs, with the same innovative design features and design methodology, regardless of the different back-gate voltage arrangements for gain control and transistor sets.

Also, unlike other works in the prior art, the power consumption of the two VG-LNAs reduces as the gain reduces, and this is a distinctive and very effective feature for energy-efficient low-power phased arrays. The comparison with the prior-art works, designed according to the well-established design approaches, shows that, irrespective of their gain states, 16-to-32 units of this new class of VG-LNAs consume less than a unit of the prior-art VG-LNAs.

Therefore, the new class of VG-LNAs provides key enabling solutions for cost-effective energy-efficient low-power phase-array ICs for next-generation (5G/6G) wireless communication systems.

ACKNOWLEDGMENTS

The authors are thankful to Keysight Technologies for the support through the donation of equipment and cad tools, and to Dr. C. Kretzschmar, Dr. P. Lengo, Dr. B. Chen (GlobalFoundries) for the technology support.